\documentclass[useAMS,usenatbib]{mn2e}
\usepackage{snapshot}
\usepackage{rotating}
\usepackage{epsfig}
\usepackage{aas_macros}
\usepackage{multirow}

\title[Separating the BL Lac and Cluster X-ray Emissions in Abell 689
  with \emph{Chandra}]{Separating the BL Lac and Cluster X-ray Emissions in
 Abell 689 with \emph{Chandra}}
\author[P. A. Giles, B. J. Maughan, M. Birkinshaw, D. M. Worrall,
K. Lancaster]{P. A. Giles\thanks{E-mail:
P.Giles@bristol.ac.uk},
B. J. Maughan, M. Birkinshaw, D. M. Worrall,
K. Lancaster \\
\\
HH Wills Physics Laboratory, Tyndall Avenue, Bristol, BS8 1TL, UK}

\begin{document}

\date{Accepted 2011 August 26. Received 2011 August 4; in original form 2011 March 10}

\pubyear{2011}

\maketitle

\label{firstpage}

\begin{abstract}

We present the results of a \emph{Chandra} observation of the galaxy
cluster Abell 689 (z=0.279).  Abell 689 is one of the most luminous 
clusters detected in the ROSAT All Sky Survey (RASS), but was flagged as
possibly including significant point source contamination.  The small
PSF of the \emph{Chandra} telescope allows us to confirm this and
separate the point source from the extended cluster X-ray emission.
For the cluster we  determine a bolometric luminosity of L$_{\rm bol}$
= (3.3$\pm$0.3)$\times$10$^{\rm 44}$ erg s$^{\rm -1}$ and a
temperature of kT=5.1$^{+2.2}_{-1.3}$ keV when including a physically
motivated background model.  We compare our measured luminosity
for A689 to that quoted in the Rosat All Sky Survey (RASS) and find 
L$_{\rm 0.1 - 2.4, keV}$ = 2.8$\times$10$^{44}$ erg s$^{-1}$, a value
$\sim$10 times lower than the ROSAT measurement.  Our analysis of the
point source shows evidence for  significant pileup, with a pile-up
fraction of $\simeq$60\%. SDSS spectra and HST images lead us to the
conclusion that the point source within Abell 689 is a BL Lac object.
Using radio and optical observations  from the VLA and HST archives,
we determine $\alpha_{\rm ro}$=0.50, $\alpha_{\rm ox}$=0.77 and 
$\alpha_{\rm rx}$=0.58 for the BL Lac, which would classify it as
being of `High-energy peak BL Lac' (HBL) type.  Spectra extracted of
A689 show a hard X-ray excess at energies above 6 keV that we
interpret as inverse Compton emission from aged electrons that may
have been transported into the cluster from the BL Lac.   

\end{abstract}

\begin{keywords}
galaxies: clusters: general -- galaxies: clusters: individual: Abell 689 -- BL Lacertae objects: general
\end{keywords}

\section{Introduction}
\label{intro}
Studies of clusters of galaxies, including measurements of their 
number density and growth from the highest density perturbations 
in the early Universe, offer insight into the underlying cosmology 
\citep{2003ApJ...590...15V,2004MNRAS.353..457A}.  However, in order 
to use clusters as a cosmological probe three essential tools are 
required \citep{2010A&A...514A..80D}: (a) an efficient method to find 
clusters over a wide redshift range, (b) an observational method of 
determining the cluster mass, and (c) a method to compute the
selection function or the survey volume in which clusters are found.  
These requirements are met by large surveys with well understood 
selection criteria. Arguably the most effective method of building 
large, well defined cluster samples has been via X-ray selection. 
The high X-ray luminosities of clusters make it relatively easy to 
detect and study clusters out to high redshifts.

Many cluster samples have been constructed based upon large X-ray
surveys such as the Einstein Medium Sensitivity Survey
\citep[EMSS;][]{1990ApJS...72..567G} and the ROSAT All Sky Survey
\citep[RASS;][]{1992eocm.rept....9V}.  However due to the relatively
poor angular resolution of these X-ray observatories, 
observations of clusters were susceptible to point source
contamination.  Indeed, within the ROSAT Brightest Cluster Sample
\citep[BCS;][]{1998MNRAS.301..881E} and its low-flux extension
\citep[eBCS;][]{2000MNRAS.318..333E}, 9 out of 201 clusters and 8 out
of 99 clusters respectively were flagged as probably having a
significant fraction of the quoted flux from embedded point sources.  
Being able to resolve these point sources is of crucial importance for
the reliable estimation of cluster properties, and indeed the nature
of the point source contamination is of independent interest.

The study of galaxy clusters has been transformed with the launch of
powerful X-ray telescopes such as \emph{Chandra} and XMM Newton, which 
have allowed the study of the X-ray emitting intracluster medium (ICM)
with unprecedented detail and accuracy.  With the launch of this new
generation of X-ray telescopes, we are able to uncover interesting
features in the morphologies of individual clusters.  In particular,
\emph{Chandra's} high angular resolution provides the means to examine
individual cluster features with great detail.       

Abell 689 \citep[hereafter A689;][]{1958ApJS....3..211A}, at z=0.279
\citep{1995MNRAS.274.1071C}, was detected in the RASS in an
accumulated exposure time of 317s.  It is included in the BCS, with a
measured X-ray luminosity of 3.0$\times$10$^{45}$ erg s$^{-1}$ 
in the 0.1--2.4 keV band.  This luminosity is the third
highest in the BCS, and thus A689 meets the selection criteria for
various highly X-ray luminous cluster samples
\citep[e.g.][]{2002ApJS..139..313D}.  However this cluster was noted
as having possible point source contamination, and for this reason has 
often been rejected from many flux limited samples.  In this study we 
present results of \emph{Chandra} observations designed to separate
any point sources and determine uncontaminated cluster properties.  

The outline of this paper is as follows.  In $\S$~\ref{reduction} we
discuss the observation and data analysis.  Results of the X-ray
cluster analysis is presented in $\S$~\ref{xray analysis}. In
$\S$~\ref{point source} we present our analysis of the X-ray point
source through X-ray, optical and radio observations.  We interpret
our results in $\S$~\ref{discussion} and the conclusions are presented
in $\S$~\ref{conclusions}.  Throughout this paper we adopt a cosmology
with $\Omega_M$= 0.3, $\Omega_{\Lambda}$ = 0.7 and 
H$_0$ = 70 km s$^{-1}$ Mpc$^{-1}$, so that 1$^{\prime\prime}$
corresponds to 4.22 kpc at the redshift of A689.  We define 
spectral index, $\alpha$, in the sense S $\propto$ $\nu^{-\alpha}$.

\section{Observations and data reduction}
\label{reduction}
The \emph{Chandra} observation of A689 (ObsID 10415) was carried out January
01, 2009.  A summary of the cluster's properties is 
presented in Table~\ref{table:a689}.  The observation was taken in 
VFAINT mode, and the source was observed in an ACIS-I configuration at
the aim point of the I3 chip, with the ACIS S2 chip also turned on.  

\begin{table*}
\begin{tabular}{cccccc}
\hline\hline
Name & RA & Dec & z & N$_{\rm H,Gal}$ & RASS L$_{ \rm {X,0.1-2.4 keV}}$ \\
\hline
Abell 689 & 08$^{\rm h}$ 37$^{\rm m}$ 24$^{\rm s}$.70 & +14$^{\rm o}$ 58$^{\prime}$ 20$^{\prime\prime}$.78 & 0.279 & 3.66$\times$10$^{20}$
cm$^{-2}$ & 30.4$\times$10$^{44}$ erg s$^{-1}$ \\
\hline
\end{tabular}
\caption{\small{Basic properties. Columns: (1) = Source name; (2) =
    Right Ascension at J2000 from {\em Chandra}; (3) = Declination at 
    J2000 from {\em Chandra}; (4) = Redshift; (5) = Galactic column 
    density; (6) = Intrinsic X-ray luminosity in the 0.1--2.4 keV band 
    based upon ROSAT observations 
\citep{1998MNRAS.301..881E}.}}
\label{table:a689}
\end{table*}

For the imaging analysis of the cluster we used the {\scriptsize CIAO}\footnote{See http://cxc.harvard.edu/ciao/} 4.2 software package with {\scriptsize CALDB}\footnote{See http://cxc.harvard.edu/caldb/} version 4.3.0 and 
followed standard reduction methods.  Since our observation was
telemetered in VFAINT mode additional background screening was applied 
by removing events with significantly positive pixels at the border 
of the 5$\times$5 event island\footnote{See
  http://cxc.harvard.edu/ciao/why/aciscleanvf.html}.  We inspected
background light curves of the observation following the 
recommendations given in \cite{2003ApJ...583...70M}, to search for 
possible background fluctuations.  The light curve was cleaned by
3$\sigma$ clipping and periods with count rates $>$20$\%$ above the
mean rate were rejected.  
curve with rejected periods showed in red.  The final level-2 event 
file had a total cleaned exposure time of 13.862 ks.

As discussed in Sect. 4, there is a bright point source at the center
of the extended cluster emission which is affected by pileup.  For the
analysis of this source we followed the same reduction method,
with the exception that VFAINT cleaning was not applied.  Applying
VFAINT cleaning leads to incorrect rejection of piled-up events,
introducing artifacts in the data.    

\begin{figure}
\begin{center}
\includegraphics[width=8.4cm]{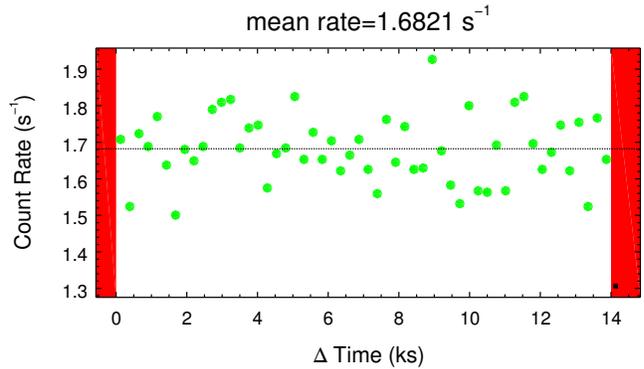}
\end{center}
\caption{\small{Background light curve for the observation of A689 in
    the 0.3-12.0 keV band. The CCD on which the cluster 
    falls (ACIS-I3) and all point sources are excluded.  The red bands show 
    periods excluded by the Good Time Interval (GTI) file.}}
\label{fig:lc}
\end{figure}

\section{X-RAY CLUSTER ANALYSIS}
\label{xray analysis}
\begin{figure*}
\begin{center}
\includegraphics[width=17.5cm]{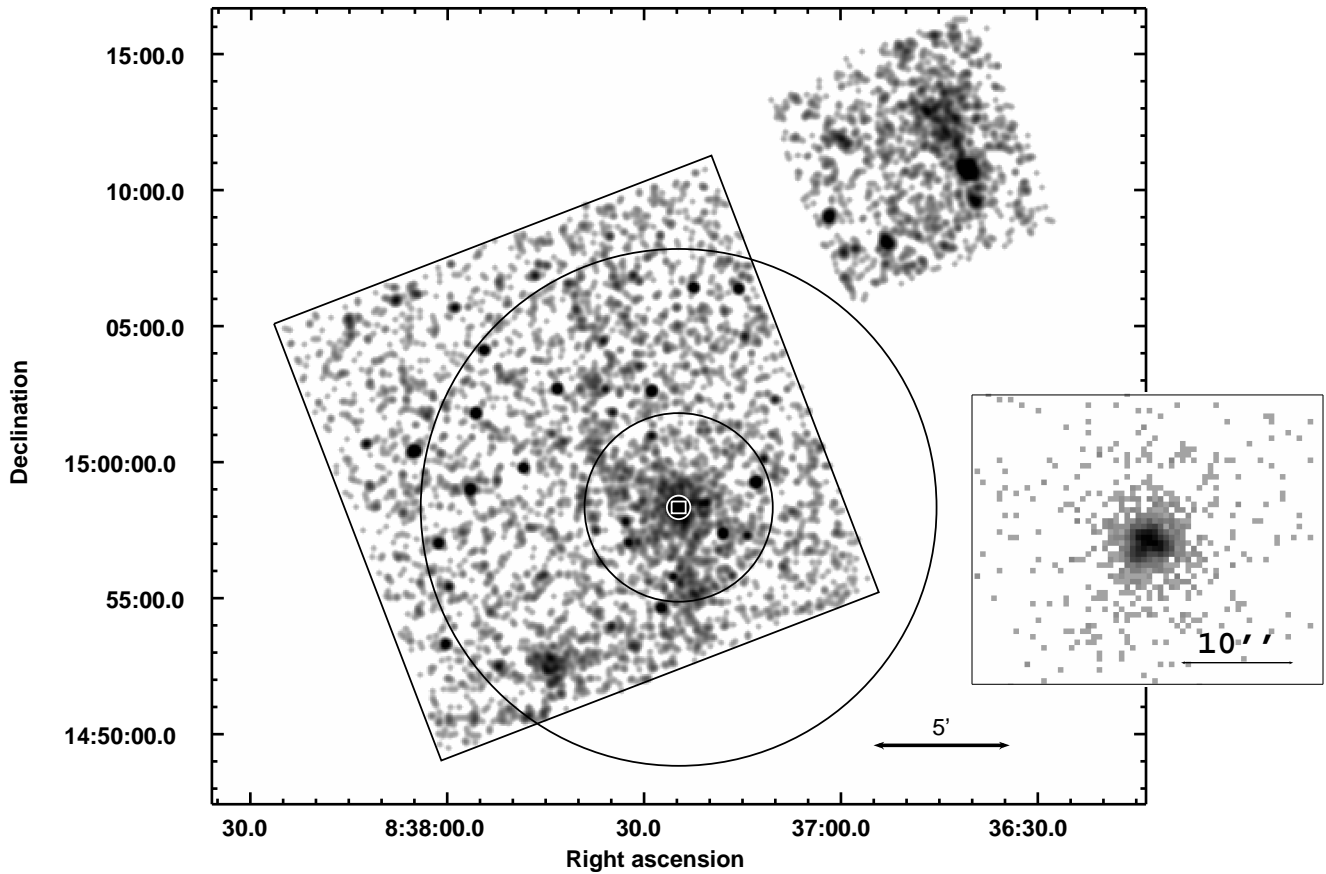}
\end{center}
\caption{\small{0.7-2.0 keV image of A689, smoothed by a Gaussian
    ($\sigma$=1.5 pixels, where 1 pixel = 3.94$^{\prime\prime}$), cleaned
    in VFAINT mode and with the readout streak removed.  Inset is a 
    zoomed-in unbinned image of the central point source within A689,
    cleaned in FAINT mode (see $\S$~\ref{reduction}).  The inner black 
    circle (r = 208$^{\prime\prime}$) is the region in which we extract the
    spectra for our analysis of the cluster emission (see
    $\S$~\ref{X-ray props}), the outer
    black circle (r = 570$^{\prime\prime}$) represents the detected
    cluster radius (see $\S$~\ref{xray analysis}), and the region
    between this and the black box was used for the local
    background (see $\S$~\ref{X-ray props}).  The white box displays
    the size of the inset, and the inner white circle 
    (r = 26$^{\prime\prime}$) shows the region excluded due to the
    central point source.  Many point sources are seen in the 
    observation and are excluded from our analysis.  An extended
    source on the NE chip can be seen, that is unrelated to A689 and 
    is also excluded from our analysis.}}
\label{fig:clust}
\end{figure*}

\begin{figure}
\begin{center}
\includegraphics[width=6.0cm, angle=270]{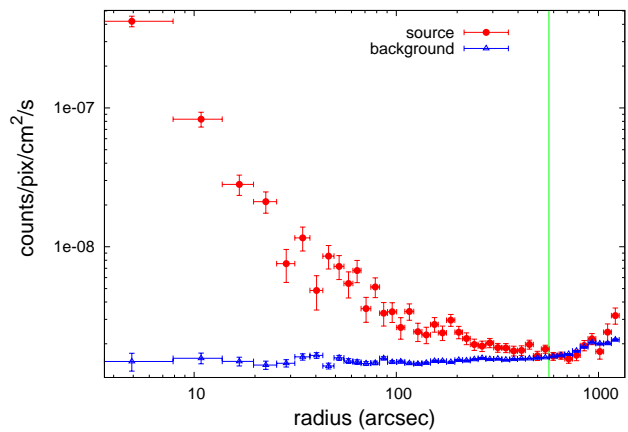}
\end{center}
\caption{\small{Exposure-corrected radial surface brightness profiles
    (0.7-2.0 keV band) of the cluster (red) and the blank-sky
    background (blue), with the green line representing the radius
    beyond which no significant cluster emission is detected.}}
\label{fig:rprof}
\end{figure}

In this section we determine global cluster properties of
Abell 689.  Figure~\ref{fig:clust} shows a Gaussian smoothed image of
the cleaned level 2 events file in the 0.7-2.0 keV band (the readout 
streak is removed using the {\scriptsize CIAO} tool {\scriptsize ACISREADCORR}), with an inset image of the point source which lies at the center of the cluster.
The extent of the diffuse cluster emission was determined by plotting 
an exposure-corrected radial surface brightness profile 
(Fig~\ref{fig:rprof}), in the 0.7-2.0 keV band, of both the 
observation and blank-sky background to determine where the cluster 
emission is lost against the background.  Figure~\ref{fig:rprof} 
demonstrates that the diffuse cluster emission is detectable to a 
radius r $\approx$ 570$^{\prime\prime}$ ($\approx$ 2.41 Mpc).  At
large radii (r$\ge$700$^{\prime\prime}$) the curves rise due to
vignetting corrections (larger at larger radii) applied to all the
counts, whereas in reality each curve contains a component from
particles that have not been focused by the telescope.    

\subsection{Background Subtraction}
\label{background}  
In order to take the background of the observation into account,
appropriate period E {\emph{Chandra} blank-sky backgrounds were 
obtained, processed identically to the cluster, and reprojected 
onto the sky to match the cluster observation.  We then followed a
method similar to that outlined in \cite{2006ApJ...640..691V}, in 
order to improve the accuracy of the background by applying small 
adjustments to the baseline model.  Firstly we corrected for the rate 
of charged particle events, which has a secular and short-term
variation by as much as 30\%.  We renormalise the background in the 
9.5--12 keV band, where the \emph{Chandra} effective area is nearly 
zero and the observed flux is due entirely to the particle background 
events.  The renormalisation factor was derived by taking the ratio of 
the observed count rate in the source and background observations 
respectively.  The normalised spectrum of the blank-sky background is 
shown in  Figure~\ref{fig:bgs}, over-plotted on the local background 
for comparison.  The spectra agree well in the 9.5--12.0 keV band, and
across the whole spectrum with only slight differences.  In addition 
to the particle background, the blank-sky and source observations 
contain differing contributions from the soft X-ray background, 
containing a mixture of the Galactic and geocoronal backgrounds, 
significant at energies $\leq$1 keV.  To take into account any 
difference in this background component between the blank-sky and 
source observations, the spectra were subtracted and residuals were
modeled in the 0.4-1keV band using an APEC thermal plasma model 
\citep{2001ApJ...556L..91S}. This model was included in the spectral 
fitting for the cluster analysis.  As can be seen in
Figure~\ref{fig:bgs} this component is very weak in the case of A689.  

\begin{figure}
\begin{center}
\includegraphics[width=6.0cm, angle=270]{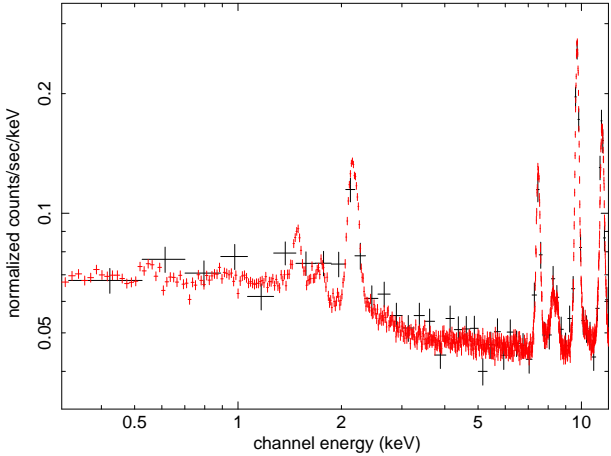}
\end{center}
\caption{\small{Comparison of the local (black) and blank-sky 
(red) background spectra, normalised to match in the 9.5-12.0 keV band.}}
\label{fig:bgs}
\end{figure}

\subsection{X-ray Cluster Properties}
\label{X-ray props}
The analysis of the diffuse X-ray emission allows us to determine 
the X-ray environment surrounding the cluster central point source.  
Throughout this process we excluded the central point source
(Fig~\ref{fig:clust}; r$\leq$26$^{\prime\prime}$) and the associated  
readout streaks to avoid contaminating the cluster emission.  

To determine cluster properties we extract spectra out to a radius
chosen so that the cluster has the maximum possible signal-to-noise 
(SNR).  The net number of counts, corrected for background, is then 
414 (with SNR = 15) and the extraction annulus is
26$^{\prime\prime}$$<$r$<$208$^{\prime\prime}$ (see
Fig~\ref{fig:clust}), centered on the cluster (at $\alpha$, $\delta$ 
= 08$^{\rm h}$ 37$^{\rm m}$ 24$^{\rm s}$.70, 14$^{\rm o}$ 58$^{\prime}$ 
20$^{\prime\prime}$.78).  We fitted the extracted spectrum in {\scriptsize XSPEC} with an absorbed thermal plasma model (WABS$\times$APEC) and 
subtracted the background described in $\S$~\ref{background}.  We 
obtain a temperature of 13.6$^{+13.2}_{-5.1}$ keV and a bolometric 
luminosity of L$_{\rm bol}$=(10.2$\pm$2.9)$\times$$10^{44}$ ergs
s$^{-1}$.  The measured temperature is far above what we would expect 
from the luminosity.  Figure~\ref{fig:LT} shows the 
luminosity-temperature relation for a sample of 115 galaxy clusters 
\citep{2008ApJS..174..117M}, along with the luminosity and temperature 
derived for A689 from our values above (pink square).  As the
\cite{2008ApJS..174..117M} sample of clusters covers a wide redshift
range, the luminosities of the clusters were corrected for the expected
self-similar evolution, given by L$_{\rm X}$$\times$E(z)$^{-1}$, where:
\begin{equation}
E(z) = \Omega_{\rm M0} (1 + z)^3 + (1 - \Omega_{\rm M0} - \Omega_{\Lambda})(1 + z)^2
+ \Omega_{\Lambda}.
\end{equation}

The same correction was also applied to the A689 data for the plot.  
Our luminosity was derived within the same annular region as the
cluster temperature and extrapolated both inward and outward in radius 
between (0-1)r$_{\rm 500}$ (where r$_{500}$ represents the radius at
which the density of the cluster is 500 times the critical density of 
the Universe at that redshift) using parameters from a $\beta$-profile 
fit to the surface brightness profile.  This takes into account our
exclusion of the region near the central point source within the
cluster, and the extrapolation to r$_{\rm 500}$ is in order to compare
with the \cite{2008ApJS..174..117M} sample.  Our r$_{\rm 500}$ value
was determined from the temperature and using the relation between
r$_{500}$ and T given in \cite{2006ApJ...640..691V}.      

The high temperature for A689 could be an indication that our observation 
suffers from background flaring which would lead to an overestimate of 
the cluster temperature.  However, no evidence is found that the
spectrum of the blank-sky background differs from that of the local 
background (Figure~\ref{fig:bgs}), or that the background of the 
observation suffers from periods of flaring (Figure~\ref{fig:lc}).  
To investigate the sensitivity of the temperature to our choice of 
background, we repeated the analysis using a local background region
far from the cluster emission (see Fig~\ref{fig:clust}).  We obtained
a temperature of 10.0$^{+13.8}_{-3.3}$ keV.  This temperature is again 
anomalous given the luminosity-temperature relation 
(Fig~\ref{fig:LT}, green triangle.  The luminosity is derived 
using the same method as above, only this time using this new value 
for the temperature to derive r$_{\rm 500}$.  The spectrum with a
local background subtracted is shown in Figure~\ref{fig:clustfit}.  We 
see an excess of photons in the $\sim$6-9 keV band, which might have 
an effect on the spectral fit.  We perform the same fit using a local 
background, however this time fitting in the 0.6-6.0 keV band to
ignore these excess high energy photons.  We obtain a temperature of 
9.4$^{+8.7}_{-2.9}$ keV.      

Figure~\ref{fig:LT} shows that the temperature is high for the
luminosity, or the cluster is X-ray under-luminous.  Before
considering a physical interpretation for the apparently high cluster
temperature, we investigated possible systematic effects from the 
background subtraction. This was done by independently modeling the 
background.              

\begin{figure}
\begin{center}
\includegraphics[width=6.0cm, angle=270]{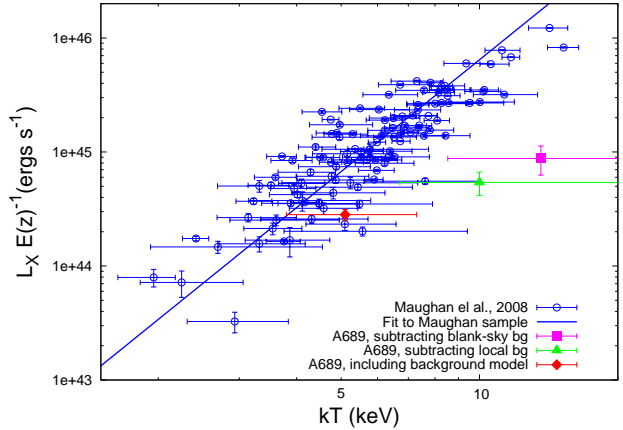}
\end{center}
\caption[]{\small{Luminosity-Temperature relation of a sample of 115
    clusters of \cite{2008ApJS..174..117M}(blue open circles).  The luminosities are measured within [0 $<$ r $<$ 1]r$_{500}$ and the temperatures within [0.15 $<$ r $<$ 1.0]r$_{500}$, in order to minimize the effect of cool cores on the derived cluster temperature.  Our derived temperatures for A689 are overplotted for comparison (pink square, green triangle, red diamond) (see $\S$~\ref{X-ray props}).}}
\label{fig:LT}
\end{figure} 

\begin{figure}
\begin{center}
\includegraphics[width=5.5cm, angle=270]{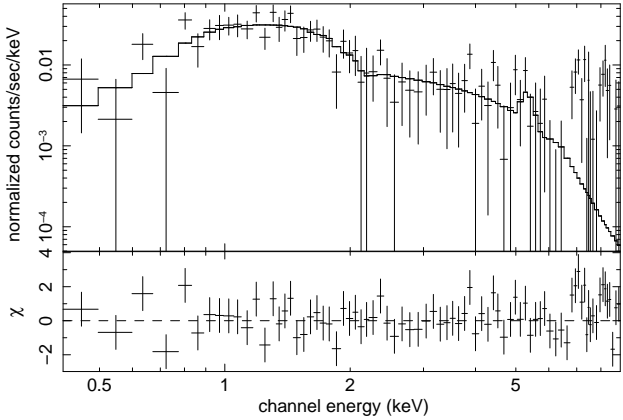}
\end{center}
\caption{\small{Spectrum of the cluster with the local background
    subtracted fitted with an absorbed thermal plasma model, with a
    reduced statistic of $\chi^2_{\rm \nu}$=1.45 ($\nu$=79).}}
\label{fig:clustfit}
\end{figure}

We model the background based upon a physical representation of its
components.  Our model consists of a thermal plasma APEC model, two 
power-law components and five Gaussian components.  The APEC model 
and one of the power-law components are convolved with files
describing the telescope and instrument response (the Auxiliary Response
File (ARF) and Redistribution Matrix File (RMF)), and are intended 
to model soft X-ray thermal emission from the Galaxy 
and unresolved X-ray background.  The second power-law component is
not convolved with the ARF as it is used to describe the high energy 
particle background, which does not vary with effective area. The five 
Gaussian components are similarly not convolved, as these are used to model
line features in the spectrum caused by the fluorescence of material 
in the telescope and focal plane caused by high energy particles.  

As described below and summarized in Table~\ref{table:model}, the
parameters of this background model were fit to the blank-sky
background, local background or high energy cluster regions or taken
from the literature, in order to build the most reliable model
possible.  We start by modeling blank-sky background in the region we
used to extract our cluster spectrum, with the model outlined
above.   This allows us to place reasonable constraints on the line 
energy and widths of each Gaussian component, and the slope of the
unconvolved power-law.  We find line features at energies of 1.48, 1.75, 
2.16, 7.48 and 8.29 keV.  We also model the convolved power-law
component, fixing the slope at a value of 1.48 taken from
\cite{2006ApJ...645...95H}.  The normalisation of this power-law
component will then be used throughout our modeling process, scaled
 by area where necessary.  A spectrum of the blank-sky background and 
a fit using our model are shown in Figure~\ref{fig:blanksky}.

Next we further constrained model parameters by fitting our
high-energy Gaussian and unconvolved power-law components in the
5.0-9.0 keV band within the cluster region. At these energies, 
the emission is dominated by particle background.  We fix the line 
energies and widths to those found in the blank-sky background and 
fit for the normalisations.  This fit finds a slope of the power-law 
consistent with that found in the blank-sky background.  We therefore 
fix the slope of the power-law at $\Gamma$=0.0061, as found for the 
blank-sky background, and fit for the normalisation.  We finally fit 
the low energy Gaussians and APEC normalisation model parameters in 
a local background region far from the cluster emission.  The high energy 
Gaussians and unconvolved power-law components are frozen at the
values found in the blank-sky and cluster regions, with the
normalisations scaled by area.  The convolved power-law 
component is frozen at the values found in the blank-sky background
and the normalisation is scaled by area.  The Gaussian features at 1.48, 1.75
and 2.16 keV are frozen at the energies and widths found in the
blank-sky background.  The temperature of the APEC model is frozen at 
0.177 keV \citep[taken from][]{2006ApJ...645...95H}. We note that our
APEC temperature is not well constrained by our data, but this is a weak 
component.  A spectrum of the local background and the corresponding 
fit with the model are shown in Figure~\ref{fig:local}.

We now model the cluster with an absorbed thermal plasma
(WABS$\times$APEC) model, including the background model outlined
above.  The normalisations of the background APEC component and the 
Gaussians at energies 1.48, 1.75 and 2.16 keV were scaled by the ratio
of the areas from the local background region to the cluster region.
The normalisations of the fluorescent lines also vary with detector
position.  To account for this effect in the low energy ($<$3 keV)
lines (where we must fit the normalisations in the off-axis local
background region), we measure the relative change in the
normalisations of each line between the local background and source
regions in the blank-sky background data.  The normalisation of each
low energy Gaussian component in the fit to the cluster data
is scaled for the different detector of the cluster region by the
relative change in normailsiation determined above (in addition to the
geometrical scaling for size of the extraction region). The
unconvolved power-law and Gaussians at 7.48 and 8.29 keV are all
frozen at the values found in the 5.0-9.0 keV cluster 
region fit.  The normalisation of the convolved power-law is frozen at
the value found in the blank-sky background.  All parameters of the
model used to describe the background are frozen in the corresponding 
cluster fit, we also freeze the redshift at 0.279 and the abundance at
0.3.  Our fit yields a temperature of 5.1$^{+2.2}_{-1.3}$ keV 
($\chi^2_{\rm \nu}$=1.15 ($\nu$=79)).  We measure a bolometric 
luminosity of L$_{\rm bol}$ = 1.7$\times$10$^{\rm 44}$ erg s$^{\rm -1}$. 
The result is shown in Figure~\ref{fig:LT} (red diamond).   The
spectrum with the corresponding fit to the cluster including the
background model is shown in Figure~\ref{fig:cluster}.  

\begin{table*}
\begin{tabular}{cc|c|c|c}
\hline\hline
Component & Represents & Parameter & Value & Where measured \\
\hline
\multicolumn{1}{|c|}{\multirow{2}{*}{Convolved power-law}} & 
\multicolumn{1}{|c|}{\multirow{2}{*}{Unresolved X-ray bg}} & 
\multicolumn{1}{c}{slope} & 1.48 &
\cite{2006ApJ...645...95H} \\ \cline{3-5}
 & \multicolumn{1}{c}{} & \multicolumn{1}{c}{normalisation} & 
4.11$\times$10$^{-6}$ & blank-sky bg \\ \cline{1-5}
\multicolumn{1}{|c|}{\multirow{2}{*}{Unconvolved power-law}} & 
\multicolumn{1}{|c|}{\multirow{2}{*}{Particle bg}} & 
\multicolumn{1}{c}{slope} & 0.061 & blank-sky bg \\ \cline{3-5}
 & \multicolumn{1}{c}{} & \multicolumn{1}{c}{normalisation} & 
0.015 & 5.0-9.0 keV cluster region \\ \cline{1-5}
\multicolumn{1}{|c|}{\multirow{3}{*}{Gaussian 1}} & 
\multicolumn{1}{|c|}{\multirow{3}{*}{Al K$\alpha$ fluorescence}} & 
\multicolumn{1}{c}{energy} & 1.48 keV & blank-sky bg \\ \cline{3-5}
 & \multicolumn{1}{c}{} & \multicolumn{1}{c}{width} & 
0.022 keV & blank-sky bg \\ \cline{3-5}
 & \multicolumn{1}{c}{} & \multicolumn{1}{c}{normalisation} & 
1.82$\times$10$^{-4}$ & local bg \\ \cline{1-5}
\multicolumn{1}{|c|}{\multirow{3}{*}{Gaussian 2}} & 
\multicolumn{1}{|c|}{\multirow{3}{*}{Si K$\alpha$ fluorescence}} & 
\multicolumn{1}{c}{energy} & 1.75 keV & blank-sky bg \\ \cline{3-5}
 & \multicolumn{1}{c}{} & \multicolumn{1}{c}{width} & 
0.95 keV & blank-sky bg \\ \cline{3-5}
 & \multicolumn{1}{c}{} & \multicolumn{1}{c}{normalisation} & 
1.45$\times$10$^{-2}$ & local bg \\ \cline{1-5}
\multicolumn{1}{|c|}{\multirow{3}{*}{Gaussian 3}} & 
\multicolumn{1}{|c|}{\multirow{3}{*}{Au M$\alpha\beta$ fluorescence}} & 
\multicolumn{1}{c}{energy} & 2.16 keV & blank-sky bg \\ \cline{3-5}
 & \multicolumn{1}{c}{} & \multicolumn{1}{c}{width} & 
0.045 keV & blank-sky bg \\ \cline{3-5}
 & \multicolumn{1}{c}{} & \multicolumn{1}{c}{normalisation} & 
2.24$\times$10$^{-3}$ & local bg \\ \cline{1-5}
\multicolumn{1}{|c|}{\multirow{3}{*}{Gaussian 4}} & 
\multicolumn{1}{|c|}{\multirow{3}{*}{Ni K$\alpha$ fluorescence}} & 
\multicolumn{1}{c}{energy} & 7.48 keV & blank-sky bg \\ \cline{3-5}
 & \multicolumn{1}{c}{} & \multicolumn{1}{c}{width} & 
0.022 keV & blank-sky bg \\ \cline{3-5}
 & \multicolumn{1}{c}{} & \multicolumn{1}{c}{normalisation} & 
5.73$\times$10$^{-3}$ & 5.0-9.0 keV cluster region \\ \cline{1-5}
\multicolumn{1}{|c|}{\multirow{3}{*}{Gaussian 5}} & 
\multicolumn{1}{|c|}{\multirow{3}{*}{Cu + Ni fluorescence}} & 
\multicolumn{1}{c}{energy} & 8.29 keV & blank-sky bg \\ \cline{3-5}
 & \multicolumn{1}{c}{} & \multicolumn{1}{c}{width} & 
0.168 keV & blank-sky bg \\ \cline{3-5}
 & \multicolumn{1}{c}{} & \multicolumn{1}{c}{normalisation} & 
4.56$\times$10$^{-3}$ & 5.0-9.0 keV cluster region \\ \cline{1-5}
\multicolumn{1}{|c|}{\multirow{4}{*}{APEC}} & 
\multicolumn{1}{|c|}{\multirow{4}{*}{Galactic foreground emission}} & 
\multicolumn{1}{c}{kT} & 0.177 keV & \cite{2006ApJ...645...95H} \\ \cline{3-5}
 & \multicolumn{1}{c}{} & \multicolumn{1}{c}{abundance} & 
1.0 & solar abundance \\ \cline{3-5}
 & \multicolumn{1}{c}{} & \multicolumn{1}{c}{redshift} & 
0 & Galactic \\ \cline{3-5}
 & \multicolumn{1}{c}{} & \multicolumn{1}{c}{normalisation} & 
2.9$\times$10$^{-5}$ & local bg \\ \cline{1-5}
\end{tabular}
\caption{\small{Table of the individual model components used to
    represent the background, with a brief interpretation of each component,
    individual component parameters, parameter values and where each
    value is calculated.  All normalisations are measured in
    photons/keV/cm$^2$/s at 1 keV, and scaled to the cluster region.
    Blank-sky background parameters were derived in the same region
    and the local background were measured in an area 2.45 times that of
    the cluster and therefore the normalisations reduced by this
    factor.  The low energy Gaussians were also corrected due to
      the normalisation dependence with position on the detector (see text).}}
\label{table:model}
\end{table*}

\begin{figure}
\begin{center}
\includegraphics[width=5.5cm, angle=270]{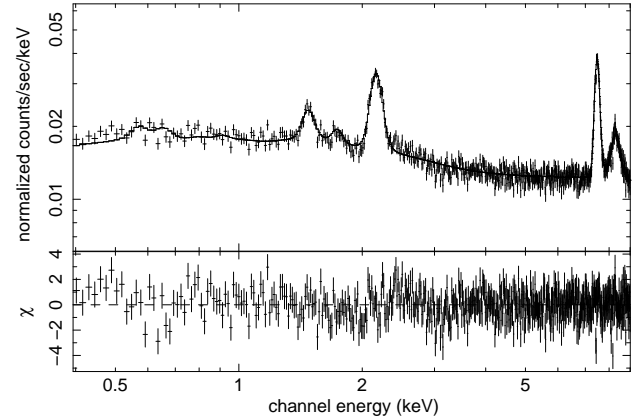}
\end{center}
\caption{\small{Spectrum of the blank-sky background in the source
    region and corresponding the fit (see Sect. 3.2), 
    $\chi^2_{\rm \nu}$=1.17 ($\nu$=585).}}
\label{fig:blanksky}
\end{figure}

\begin{figure}
\begin{center}
\includegraphics[width=5.5cm, angle=270]{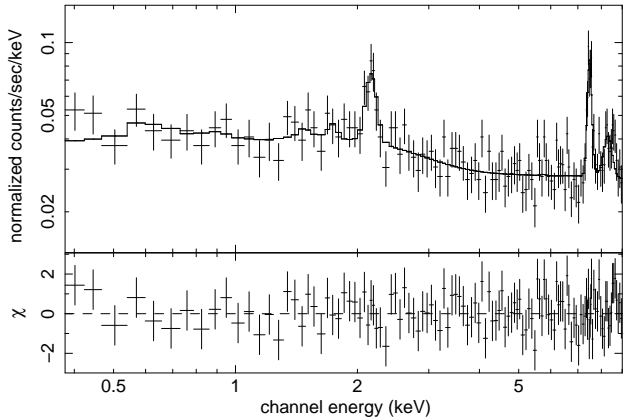}
\end{center}
\caption{\small{Spectrum of the local background and corresponding
    fit (see Sect 3.2). $\chi^2_{\rm \nu}$=0.79 ($\nu$=120).}}
\label{fig:local}
\end{figure}

\begin{figure}
\begin{center}
\includegraphics[width=5.5cm, angle=270]{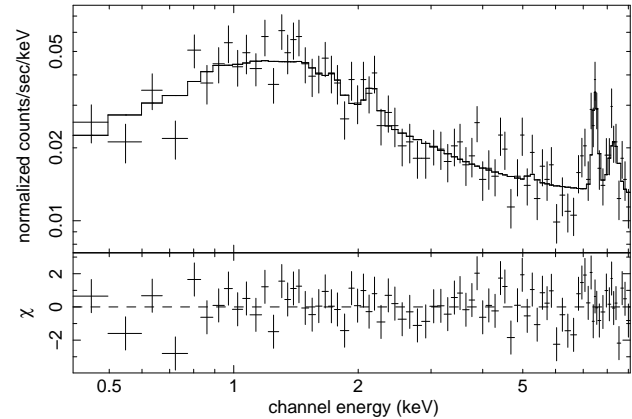}
\end{center}
\caption{\small{Spectrum of the cluster plus background fit with an 
absorbed thermal plasma model including a background model (see Sect 3.2).
$\chi^2_{\rm \nu}$=1.15 ($\nu$=79).}}
\label{fig:cluster}
\end{figure}
     
\smallskip
\section{The Central Point Source}
\label{point source}
The point source is displayed in Figure~\ref{fig:a689zoom}.  
The presence of strong readout streaks indicate that the point source is
likely to be affected by pile-up. The readout streaks occur as X-rays 
from the source are received during the ACIS parallel frame transfer, 
which provides 40$\mu$s exposure per frame in each pixel along the
streak.  We detail our analysis of the point source and estimate of the 
pileup fraction in the following section. 

\begin{figure}
\begin{center}
\includegraphics[width=8.4cm]{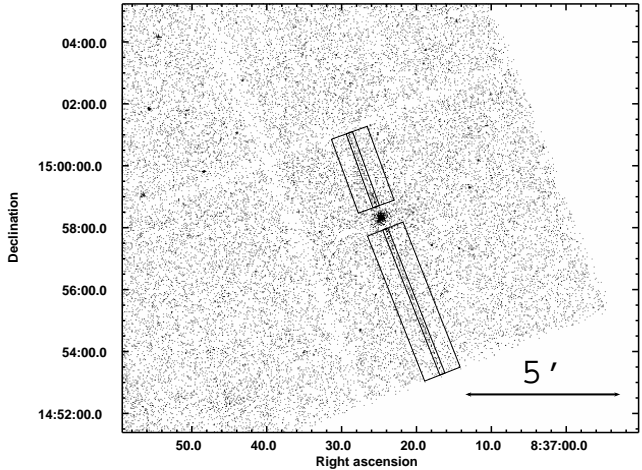}
\end{center}
\caption{\small{\emph{Chandra} image of A689 showing the regions used for
    extracting spectra of the readout streak (inner rectangles) and
    the corresponding background regions for the readout streak (outer
rectangles).}}
\label{fig:a689zoom}
\end{figure}

\subsection{X-ray Analysis of the Point Source}
\label{xray point}
As a first test of the predicted pile-up, we compared the image of the
point source to the \emph{Chandra} Point Spread Function (PSF).  We
made use of {\scriptsize CIAO} tool {\scriptsize MKPSF} to create an image of the on-axis PSF following the method outlined in \cite{2003A&A...407..503D}.  This consists of merging 7 different monochromatic PSFs chosen and weighted
on the basis of the source energy spectrum between 0.3 and 8 keV.
This method can be summarized as follows:
\begin{enumerate}
\item We first extract the energy distributions of the photons from
a circular region centered on the peak brightness of the source with a
radii of 2.5$^{\prime\prime}$.
\item We choose seven discrete energy values at which to creating each
  PSF, with the number of counts at each energy corresponding to that 
PSF's `weight'.  
\item Using {\scriptsize MKPSF} we create seven monochromatic PSFs at the position of the point source on the detector and co-added them.  Each PSF is
weighted by its relative normalisation (found in the previous step).  
\end{enumerate}

\begin{figure}
\begin{center}
\includegraphics[width=6.0cm, angle=270]{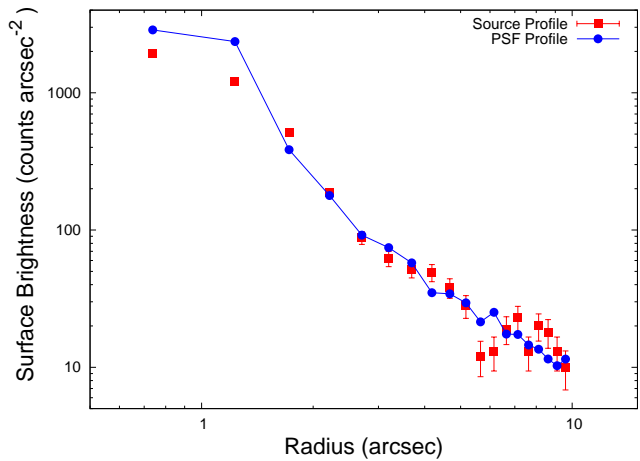}
\end{center}
\caption{\small{Surface brightness profiles of the point source (red
    squares) and the composite PSF (blue circles) in the 0.3-8.0 keV
    band, normalised to agree in the 2.46-4.92 arcsec radii region.}}
\label{fig:PSFcomp}
\end{figure}

Once we obtained the composite PSF, we normalised it to the counts 
within an annulus (inner and outer radius 2.46 and 4.92 arcsec
respectively) in order to avoid any effects of pile-up in the central
region.  We then compare surface brightness profiles of the point
source and PSF to look for evidence of pile-up in the core of the
point source image.  Figure~\ref{fig:PSFcomp} shows the radial surface
brightness profiles of the point source (red) and the composite PSF
(blue).  We find that the point source and PSF agree well in the wings 
of the PSF ($>$ 2.46$^{\prime\prime}$) and that there is an excess in the 
PSF surface brightness above that of the point source at the peak of
the source.  This is consistent with the flattening of the source
profile relative to that of the PSF due to pileup in the core.  
The PSF then gives an estimate of the non piled up count rate.  Given
this count rate we use
PIMMS\footnote{http://cxc.harvard.edu/toolkit/pimms.jsp} 
to estimate a pile-up fraction of 65\%.

We also compute a second estimate of the core count rate using the
ACIS readout streak.  By fitting a model to the spectrum extracted
from the readout streak we can compare this to a spectrum extracted in 
the core and fitted using a pileup model.  We follow the method outlined 
in \cite{2005ApJS..156...13M} in order to correct the exposure time of
the readout streak spectrum.  For an observation of live time
t$_{live}$, a section of the readout streak that is $\theta_s$ arcsec
long accumulates an exposure time of 
t$_{\rm s}$ = 4 $\times$10$^{-5}$t$_{\rm live}$$\theta_{\rm s}$/(t$_f$$\theta_{\rm x}$) s, where 
$\theta_{\rm x}$ = 0.492$^{\prime\prime}$ is the angular size of an ACIS
pixel.  For our observation, t$_{\rm live}$ = 13.862 ks and the frame time
parameter t$_{\rm f}$ = 3.1 s, giving t$_{\rm s}$ = 165s in a streak segment that is 
454$^{\prime\prime}$ long.  Figure~\ref{fig:a689zoom} shows the
regions used for extracting spectra of the readout and an adjacent
background region.  This choice of background region ensures the
cluster emission is subtracted from the readout streak spectrum.
Using the {\scriptsize SHERPA} package \citep{SciPyProceedings_51} we fitted an absorbed 1-D power-law model (WABS$\times$POWER-LAW) to the
extracted spectrum of the readout streak.  We obtain fit parameters
for the photon index = 2.33$^{+0.34}_{-0.30}$ and a normalisation of 
0.0033$\pm0.0005$ photons keV$^{-1}$ cm$^{-2}$ s$^{-1}$ 
($\chi^2_{\rm \nu}$=0.34 ($\nu$=63)).  
We then extract a spectrum of the point source in a region of radius 
5$^{\prime\prime}$, and subtracted the same background as for the
readout streak.  We once again fitted an absorbed power-law model, 
including this time a pileup model (jdpileup).  We obtain fit
parameters for the photon index = 2.22$^{+0.05}_{-0.04}$, consistent
to that found from the readout streak, and a normalisation of 
0.0015$\pm$0.0001 photons keV$^{-1}$ cm$^{-2}$ s$^{-1}$ 
($\chi^2_{\rm \nu}$=1.6 ($\nu$=119)).  The extracted spectrum and 
corresponding fit is shown in Figure~\ref{fig:pileup}.  The pileup 
fraction is estimated to be 60\%, which is consistent with that 
found using the PSF count rate.  The normalisation found in the fit
can be converted to an X-ray flux density for this source, 
f$_{\rm 1~keV}$=0.99$\pm$0.07 $\mu$Jy.           

\begin{figure}
\begin{center}
\includegraphics[width=7.0cm, angle=270]{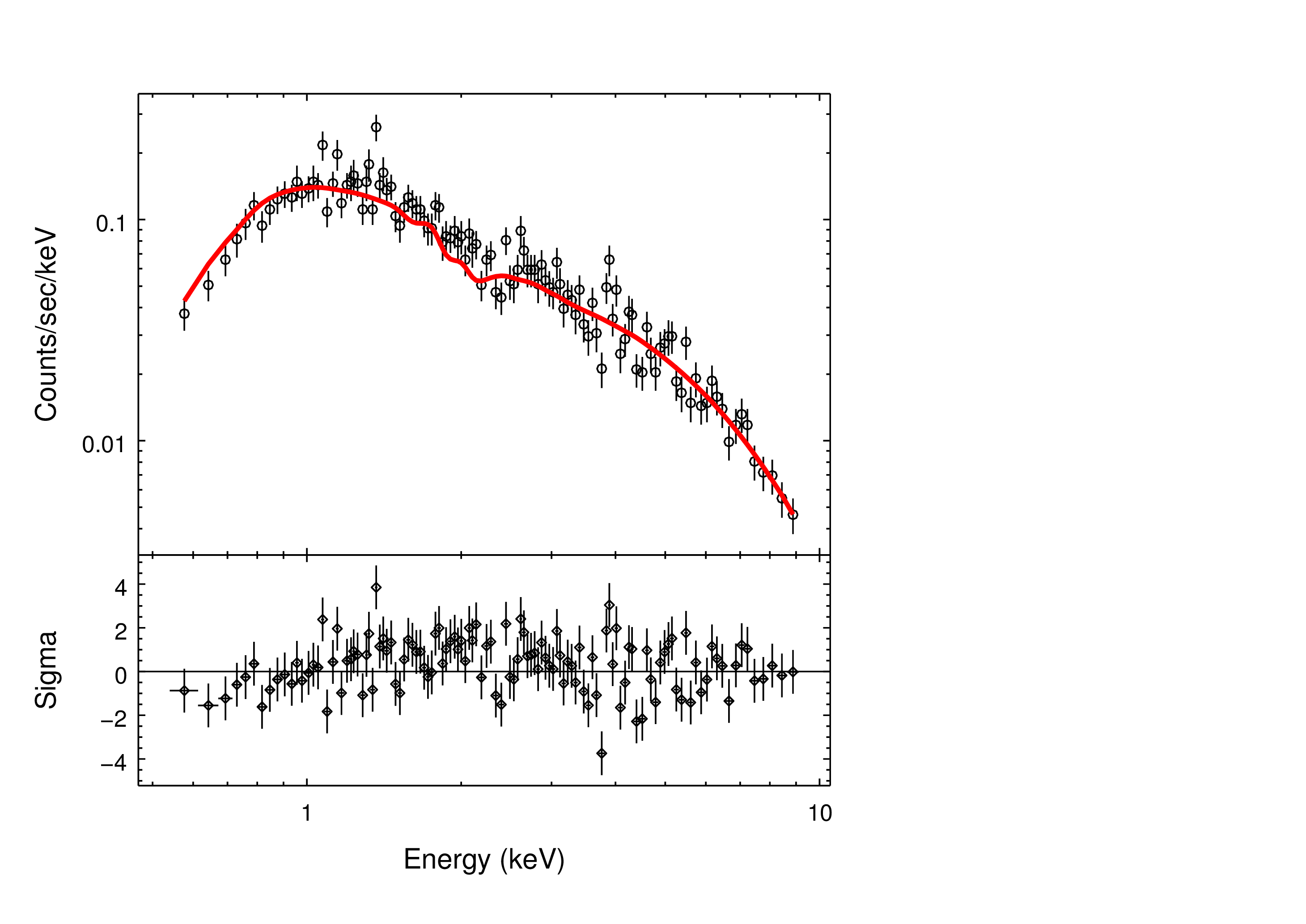} \\
\end{center}
\caption{\small{Spectra of the point source fitted with an absorbed 
power-law, including a pileup model. $\chi^2_{\rm \nu}$=1.6 ($\nu$=119).}}
\label{fig:pileup}
\end{figure}

\subsection{Optical Observations}
\label{optical point}
Abell 689 was observed with the Hubble Space Telescope (HST) with the 
F606W filter (\~V-band) on January 20, 2008.  Marking the
position of the peak X-ray emission of the point source on the HST 
image, we find that this corresponds to an object resembling an active
nucleus in a relatively bright galaxy 
(Fig~\ref{fig:hst}).  We searched the SDSS DR7 archive for information 
on the spectral properties of this object.  At the coordinates of the 
X-ray point source (SDSS coordinates of $\alpha$,$\delta$ = 
08$^{h}$ 37$^m$ 24.7, 14$^{o}$ 58$^{\prime}$ 19$^{\prime\prime}$.8) we find a 
blue object with a corresponding relatively featureless spectrum
(Fig~\ref{fig:sdss}).  From SDSS we quote an r-band magnitude of
17.18.  The spectrum resembles that of a BL Lac object, 
a type of AGN orientated such that the relativistic jet is closely 
aligned to the line of sight.  From the H and K lines in the spectrum 
(dotted green lines in Fig~\ref{fig:sdss}), thought to be from the
host galaxy, the redshift is determined to be z=0.279, consistent with
the redshift assigned to the cluster \citep{1995MNRAS.274.1071C}.
Using the HST observation we measure an optical flux for the BL Lac of 
f$_{\rm 5997\AA}$=112mJy.               

\begin{figure}
\begin{center}
\includegraphics[width=8.5cm]{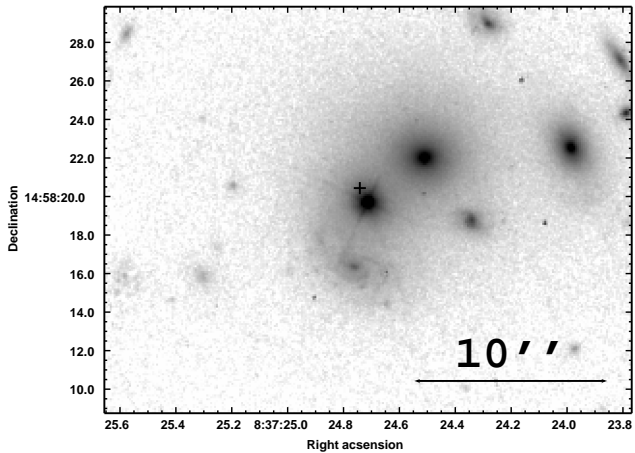}
\end{center}
\caption{\small{HST image of the point source with a cross marking the
    position of the peak of the X-ray emission.}}
\label{fig:hst}
\end{figure}

\begin{figure}
\begin{center}
\includegraphics[width=8.5cm]{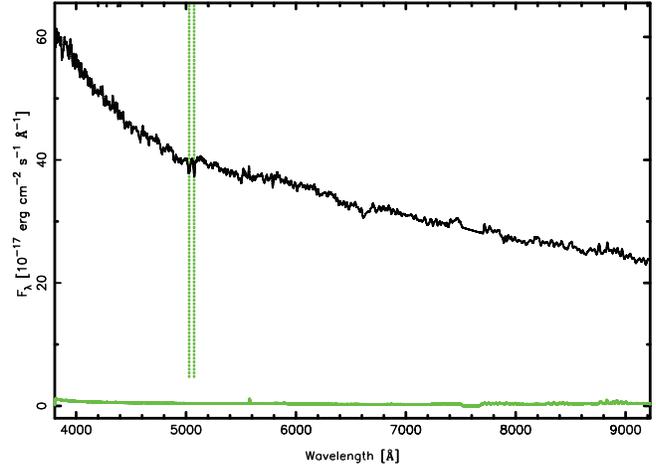}
\end{center}
\caption{\small{Spectra of the BL Lac from SDSS (SDSS
    J083724.71+145819.8).  The vertical green lines represent the H
    and K lines thought to be from the host galaxy and the green line
    at the bottom represents the error spectrum.}}
\label{fig:sdss}
\end{figure}

\subsection{Radio Observations}
\label{radio point}
Archival radio observations of Abell 689 are available, allowing us to
determine the radio spectral index, $\alpha$.  We obtained 8.46 GHz 
data taken in March 1998 from the VLA archive which we mapped using AIPS.
The source is unresolved at 8.46 GHz, and we measure a flux density 
of 18.6$\pm$0.27 mJy.  We also obtained a 1.4 GHz radio image from 
the FIRST survey, from which we measure a flux of 62.7$\pm$0.25 mJy.  
From these data we obtain a spectral index of 
$\alpha_{\rm r}$$\sim$0.67$\pm$0.01.  We note that a slight angular
extension in the FIRST survey suggests that the 8.46 GHz image may be 
missing some flux density.   

\section{Discussion}
\label{discussion}

\subsection{The ICM properties of Abell 689}

\begin{table*}
\begin{tabular}{c|ccccc}
\hline\hline 
& r$_{500}$ & T$_{X}$ & L$_{\rm X,bol}$ & & \\
Background Subtraction & (arcsec) & (keV) & ($\times$10$^{44}$ erg
s$^{-1}$) & reduced $\chi^2$ & degrees of freedom \\ 
\hline
Blank-sky & 1130 & 13.6$^{+13.2}_{-5.1}$ & 10.1$\pm$2.9 & 1.19 & 74 \\
Local & 696 & 10.0$^{+13.8}_{-3.3}$ & 6.2$\pm$1.4 & 1.45 & 79 \\
Physically motivated model$^{\dagger}$ & 266 & 5.1$^{+2.2}_{-1.3}$ &
3.3$\pm$0.3 & 1.15 & 79 \\
\hline
\end{tabular}
\caption{\small{Table listing the derived spectral properties of A689
    for each of the background treatments we employ in our
    analysis. $^{\dagger}$ indicates our favored method for determining
    the cluster properties of A689.}}
\label{tab:spec_results}
\end{table*} 

We have derived the ICM properties of A689 using three methods of
background subtraction.  Table~\ref{tab:spec_results} shows the
spectral properties of the ICM for each of the background treatments
we employ with our favored method coming from a physically motivated
model of the background components.  Through detailed modeling of the 
local and blank-sky backgrounds for A689, and including this
background model in a spectral fit to the cluster, we have determined 
a temperature and luminosity for A689 of T = 5.1$^{+2.2}_{-1.3}$ keV
and L$_{\rm bol}$ = 3.3$\times$10$^{44}$ erg s$^{-1}$.  Plotting these 
values on the luminosity-temperature plot (Fig~\ref{fig:LT}) and 
comparing to the large X-ray sample of \cite{2008ApJS..174..117M}, 
we find that A689 is observed to be at the edge of the observed
scatter in the luminosity-temperature relation.  This suggests that 
either the temperature of the ICM has been enhanced or suppression of
the luminosity has occurred.  

It has been shown that systems that host a radio source are likely to 
have higher temperatures at a given luminosity.
\cite{2005MNRAS.357..279C} showed that this is the case for radio-loud
galaxy groups, and \cite{2007MNRAS.379..260M} showed that for clusters
that host a radio source there is a departure from the typical
luminosity-temperature relation, particularly in the case of low mass
systems.   \cite{2005MNRAS.357..279C} also showed, through analysis of 
{\em Chandra} and {\em XMM-Newton} observations, evidence for
radio-source interaction with the surrounding gas for many of the 
radio-loud groups.  A similar process could be occurring within A689, 
which has a confirmed radio source at the center of the cluster.  We
note that more detailed X-ray and radio observations would to be
needed in order to test any interaction between the BL Lac and ICM.  

The other possible explanation for the offset of radio-loud systems
from the luminosity-temperature relation is the suppression of the 
luminosity, which could be caused by displacement of large amounts of gas 
due to the interaction of the radio source with the ICM.  The
interaction of the radio source with the ICM will cause an increase in the 
entropy of the local ICM. This higher entropy gas will be displaced 
so that it is in entropy equilibrium with the surrounding gas.  
However, \cite{2007MNRAS.379..260M} showed that there is a correlation 
between the radio luminosity and the heat input required to produce
the observed temperature increment in clusters hosting radio sources.
They note that this correlation favors an enhanced temperature
scenario caused by the radio galaxy induced heating.          

\subsection{Comparison with the BCS}

A689 was noted in the BCS as having a significant fraction of its
flux coming from embedded point sources.  We have confirmed that A689
contains a point source at the center of the cluster and is that of a
BL Lac object.  We stated that the measured X-ray luminosity for A689 
as quoted in the BCS (L$_{\rm 0.1 - 2.4 keV}$ = 3$\times$10$^{45}$ erg
s$^{-1}$, see $\S$~\ref{intro}), is the third brightest in the BCS.
From our follow up observation with Chandra we calculate the
luminosity and compare with the BCS value.  The luminosities in the
BCS are calculated within a standard radius of 1.43~Mpc, which at the 
redshift of A689 corresponds to a radius of 338$^{\prime\prime}$.  We 
therefore employ the same method as in Sect~\ref{X-ray props} and 
integrate under a beta model fitted to the derived surface brightness 
profile and extrapolate inward and outward from
26-206$^{\prime\prime}$ to 0-338$^{\prime\prime}$.  The unabsorbed
flux in the 0.1 - 2.4 keV band (observed frame) was f$_{\rm 0.1-2.4, keV}$
= 5.8$\times$10$^{-13}$ erg s$^{-1}$ cm$^{-2}$. After k-correction the 
X-ray luminosity in the 0.1 - 2.4 keV band (rest frame) was 
L$_{\rm 0.1 - 2.4, keV}$ = 2.8$\times$10$^{44}$ erg s$^{-1}$.  Note
that we assume an H$_{0}$ of 50 for consistency with the BCS catalog.  
This value is $\sim$10 times lower than that quoted in the BCS, and 
A689 is now ranked 110th out of 201 in luminosity.               

\subsection{Classifying the BL Lac}

BL Lac objects may be split into `High-energy peak BL
Lacs' (HBL) and `Low-energy peak BL Lacs' (LBL), for objects which
emit most of their synchrotron power at high (UV--soft-X-ray) or 
low (far-IR, near-IR) frequencies respectively
\citep{1995MNRAS.277.1477P}.  HBL and LBL objects 
have radio-to-X-ray spectral indies of $\alpha_{rx}$$\le$0.75 and  
$\alpha_{rx}$$\ge$0.75, respectively.  We calculate a radio-to-X-ray 
spectral index for the BL Lac in A689 of $\alpha_{rx}$=0.58$\pm$0.04.  
From this value we classify our BL Lac as an HBL type.  We also
compared our BL Lac with those of \cite{1998MNRAS.299..433F}, who 
investigated the properties of large samples of BL lacs at radio to 
$\gamma$-ray wavelengths.  Our value of $\alpha_{rx}$=0.58 falls into 
the region 0.35$\le$$\alpha_{rx}$$\le$0.7, dominated by X-ray selected 
BL Lacs (XBL).  This is consistent with the X-ray selection of this 
cluster.  Using the measured HST flux, we also calculate 
$\alpha_{ro}$=0.50 and $\alpha_{ox}$=0.77.  These values are not
atypical for BL Lac objects \citep[Figure 7 in][]{1999ApJ...516..163W}. 

\cite{2003A&A...407..503D} tried to determine whether HBLs and LBLs
were characterized by different environments.  They found that of 
5 sources exhibiting diffuse X-ray emission that 4 were HBLs and 1 
was an LBL.  The BL Lac in A689 continues this trend, as it appears 
to be an HBL embedded within a cluster environment.       

\subsection{Evidence for Inverse-Compton Emission}

Our initial analysis of the extended emission in A689 yielded temperature
estimates that were significantly higher than expected based upon its
X-ray luminosity.  We also found evidence for an hard excess of X-ray 
photons in the 6.0-9.0 keV band of the cluster spectrum.  Before
assigning a physical cause we must be careful to eliminate systematic 
effects in the background subtraction, as underestimating the particle
background will give a hard excess.  This is unlikely to be the case
here as a high temperature was obtained when either a blank-sky and
local background was used.    

When using our background model instead of subtracting a background
spectrum, the cluster temperature was more consistent with its
luminosity (Fig~\ref{fig:LT}).  This appears to be due to the 
unconvolved power law fitting a hard excess in the cluster region.  
In order to assess what impact the high-energy particle component has 
on the cluster temperature, we varied the normalisation of the 
unconvolved power-law component of our background model (sect. 3.2) 
within its 1$\sigma$ errors.  The temperature ranged from 3.86 to 8.51 
keV.  Thus small changes in the particle background have a significant 
influence on the cluster temperature.  Our power-law component was 
derived within the cluster region in the 5.0-9.0 keV band, so our 
background model may be removing a physical hard X-ray component 
associated with the cluster.  We assess the possible properties of
such a component by re-deriving the normalisation of the unconvolved 
power-law component in the local background region, scaling this value 
by the ratio of the areas, and using this value for the unconvolved 
power-law normalisation in the background model for our cluster fit.  
We find a power-law normalisation of 0.0129
photons keV$^{-1}$ cm$^{-2}$ s$^{-1}$ (as opposed to 0.0149
photons keV$^{-1}$ cm$^{-2}$ s$^{-1}$ in the cluster region). Using this value and 
re-fitting (Fig~\ref{fig:plcomp}), we find a cluster temperature of 
14.3$^{+15.6}_{-5.4}$ keV.  This suggests that the hard component is 
spatially associated with the cluster.  

\begin{figure}
\begin{center}
\includegraphics[width=5.5cm, angle=270]{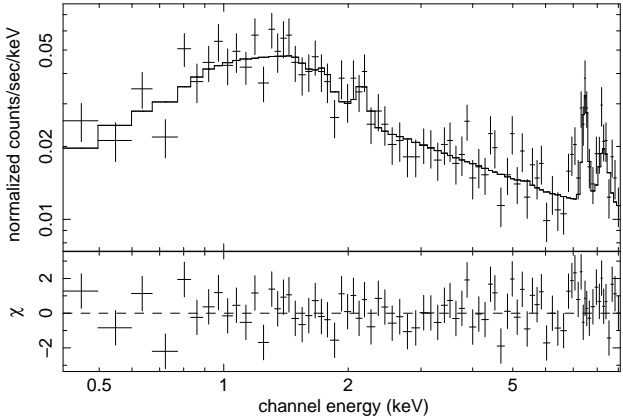}
\end{center}
\caption{\small{Spectrum of the cluster including our physically
    motivated model, with the normalisation of the unconvolved 
    power-law component derived in the local background region. 
    $\chi^2_{\rm \nu}$=1.17 ($\nu$=79)}}
\label{fig:plcomp}
\end{figure}

\cite{2007ApJ...668..796B} find a similar excess of X-ray emission in 
the cluster A3112, which is known to have a central AGN.  It is argued 
that this excess may be due to emission of a non-thermal component.  
Relativistic electrons in the intergalactic medium will cause CMB 
photons to scatter into the X-ray band (inverse Compton scattering).  
The same process could occur in A689, with the relativistic particles 
responsible for the inverse Compton scattering provided by the jets of 
the BL Lac.

In order to test the assumption of inverse-Compton emission, we add a
convolved power-law component with $\alpha$=1.5  (appropriate for
modeling IC emission from aged electrons).  We fit for the
normalisation of this added power-law component and freeze all other
parameters of the background and cluster model.  We note that we use
the normalisation of the unconvolved power-law found in the local
background region as found above (a value of 0.0129).  When fit, the
$\chi$$^2$ increases slightly but the fit is still acceptable at the
95\% confidence level.  From the additional power-law component we measure 
a 1 keV flux density of $\sim$7 nJy.  If the X-ray emission is from 
scattering of the CMB by an aged population of electrons of power-law 
number index 4.0, we can determine what the implied synchrotron
emission would be at 1.4 GHz, given plausible magnetic fields.  
Clusters typically have magnetic fields of a few $\mu$G 
\citep{2002ARA&A..40..319C}.  The NVSS survey would have detected a
flux density of $\sim$10 mJy over the entire cluster.  Adopting a 
1.4 GHz flux density limit of 10 mJy, we require a magnetic field of 
B $\leq$ 2 $\mu$G over the cluster to avoid over-predicting radio
emission.  We also calculate a value for the minimum-energy magnetic
field, B$_{\rm E\_min}$, that would give a 1.4 GHz flux density of
S$_{\rm 1.4~GHz}$ = 10~mJy, which in this case is B$_{\rm E\_min}$ = 7.5
$\mu$G.  For a flux density of S$_{\rm 1.4~GHz}$ = 5 mJy, we would require
a magnetic field of 1.6~$\mu$G and B$_{\rm E\_min}$ = 6.5~$\mu$G.  Note that 
B $\propto$ S$_{\rm 1.4~GHz}$$^{1 / 1 + \alpha}$ and B$_{\rm E\_min}$
$\propto$ S$_{\rm 1.4~GHz}$$^{1 / 3 + \alpha}$.  We conclude that the
excess X-ray emission can be attributed to inverse-Compton scattering 
without over-predicting the radio emission if the magnetic field
strength is in a range typical of clusters and within a factor of a
few of the minimum energy value.

\section{Conclusions}
\label{conclusions}
We have used a 14 ks \emph{Chandra} observation of the galaxy cluster A689 in
order to determine the nature of the cluster's point source
contamination and to analyze the cluster properties excluding the
central point source.  Our main conclusions are as follows.
\newcounter{qcounter}
\begin{list}{ \arabic{qcounter}.~}{\usecounter{qcounter}}
\item Using background subtraction of both local and blank-sky
  backgrounds, we obtain temperatures which are high relative to the
  luminosity-temperature relation.
\item We construct a physically motivated model for the background and 
  include this model in a fit to the cluster spectrum.  If the
  particle background in the cluster is allowed to exceed that in the
  local and blank-sky backgrounds we obtain a temperature
  of 5.1$^{+2.2}_{-1.3}$ keV.  However, there is no reason for there to
  be a higher particle rate in the specific region of the CCD in which
  the cluster lies.  A hard excess needed to bring the temperature to
  a reasonable value must have a different origin. 
\item We confirm the presence of a point source within A689 as
  suspected in the BCS.  When excluding the point source and using
  our derive background model we find a luminosity of 
  L$_{\rm 0.1 - 2.4, keV}$ = 2.8$\times$10$^{44}$ erg s$^{-1}$, a
  value $\sim$10 times lower than quoted in the BCS. 
\item From the X-ray analysis of the point source we find a ``flat-topped'' 
point source with a pileup fraction of $\simeq$ 60\%.
\item Optical observations of the cluster from SDSS and HST lead us
to conclude that the point source is a BL Lac type AGN.
\item We classify the BL Lac as an `High-energy peak BL Lac' with 
$\alpha$$_{rx}$=0.58$\pm$0.04.
\item We interpret the hard X-ray excess needed to bring the cluster
  temperature to a reasonable value as inverse-Compton emission from
  aged electrons that may have been transported into the cluster from
  the BL Lac.
\end{list}

We have shown here not only the importance of resolving and excluding 
point sources in cluster observations, but also the effect these point
sources can have when determining the ICM properties of galaxy clusters.
The detailed analysis we have performed here may not however be
suitable for all clusters as it is unclear whether this analysis can
be performed at higher redshifts.  Separating the point source and
cluster emissions becomes increasingly difficult at high redshifts,
however Chandra has proved capable at resolving point source emission
in clusters to z$\sim$2 \citep[e.g.][]{2009A&A...507..147A}.
Resolving point source and cluster emission out to high redshift
becomes more important with redshift due to the evolution of
the number density of point sources within clusters 
\citep{2009ApJ...694.1309G}.  The area used to model the background
components associated with the high energy cluster regions will
decrease spatially with redshift, and separating the between thermal and
inverse Compton emissions at higher redshift will become increasingly
difficult \citep{2003MNRAS.341..729F}, due to the increase in the
energy density of the CMB with redshift.

\section*{Acknowledgments}

We thank J.Price for useful discussions regarding the SDSS and Hubble
data.  We thank Ewan O'Sullivan and Dominique Sluse for useful
discussions on the nature of the point source.  We thank the anonymous
referee for valuable comments and suggestions.  PG also acknowledges 
support from the UK Science and Technology Facilities Council.  



\bsp

\label{lastpage}

\end{document}